\documentclass[prd,eqsecnum,floatfix]{revtex4}%
\usepackage{amsmath}
\usepackage{graphicx}
\usepackage{amsfonts}
\usepackage{amssymb}%
\providecommand{\U}[1]{\protect\rule{.1in}{.1in}}
\newtheorem{theorem}{Theorem}
\newtheorem{acknowledgement}[theorem]{Acknowledgement}

\newcommand{\BOX}{\hbox {$\sqcap$ \kern -1em $\sqcup$}}

\newcommand{\be}{\begin{equation}}
\newcommand{\ee}{\end{equation}}
\newcommand{\ba}{\begin{eqnarray}}
\newcommand{\ea}{\end{eqnarray}}
\newcommand{\ban}{\begin{eqnarray*}}
\newcommand{\bea}{\begin{eqnarray}}
\newcommand{\eea}{\end{eqnarray}}
\newcommand{\ean}{\end{eqnarray*}}
\newcommand{\barr}{\begin{array}}
\newcommand{\earr}{\end{array}}

\begin{document}
\title{The Omega Effect as a Discriminant for Space-Time Foam}
\author{Sarben Sarkar}
\affiliation{King's College London, University of London, Department of Physics, Strand,
London WC2R 2LS, U.K.}
\affiliation{}

\begin{abstract}
{\small If there is CPT violation, the nature of entanglement for neutral meson pairs produced in meson factories may, on general grounds, be affected. The new form of entanglement is the omega effect. Gravitational decoherence, due to
space-time foam, may be one route for deviations from CPT invariance. Two
models of space-time foam are considered. One, based on non-critical string
theory, is able to produce the new correlations in a natural way. The other,
based on the paradigm of thermal-like baths, is shown to be surprisingly
resistant to producing the effect even on exercising a total freedom of choice
for the state of the bath.}

\end{abstract}
\maketitle


\section{Introduction}

The operator $\Theta=CPT$ \ where $C$ is the charge conjugation operator, $P$
is the parity operator and $T$ the time reversal operator is a symmetry of the
S matrix for local, unitary and Lorentz invariant field theories
\cite{streater}. If one of these assumptions is not valid then $\Theta$ may
cease to be a symmetry or be ill defined. We shall consider the situation
where an otherwise relativistic theory is not unitary. The lack of unitarity
at a fundamental level will be postulated as due to space-time foam
\cite{wheeler}. In the decoherence scenario the S-matrix of the effective
low-energy field theory would then have to be replaced by a linear
non-factorisable superscattering operator $\not S  $ relating initial and
final-state density matrices $\rho$~\cite{nonunitary}
\begin{equation}
\rho_{out}=\not S  \rho_{in}.
\end{equation}
If this is correct then the usual formulation of quantum mechanics has to be
modified; from the theory of open systems (see e.g. \cite{gardiner}%
,\cite{sarkar2}) we know that, on tracing over the environmental degrees of
freedom, a system can be described by a master equation
\begin{equation}
\partial_{t}\rho=\frac{i}{\hbar}\left[  \rho,H\right]  +\Lambda\rho
\label{liouvqm}%
\end{equation}
where $\Lambda$ is a Liovillian superoperator. The foam can be regarded as an
environment and so the open systems point of view is also a natural one. It
should be pointed out that this suggestion may be invalid if there is
\emph{holography} in quantum gravity~\cite{thooft}, such that any information
on quantum numbers of matter, that at first sight appears to be lost into the
horizon, is somehow reflected back from the horizon surface, thereby
maintaining quantum coherence. This may happen, for instance, in some highly
supersymmetric effective theories of strings \cite{maldacena}, which however
do not represent realistic low-energy theories of quantum gravity.
Supersymmetry breaking complicates the issue, thus spoiling complete
holography. Recently, S. Hawking, inspired by the above recent ideas in string
theory, has also argued against the loss of coherence in a \emph{Euclidean}
quantum theory of gravity. In such a model, summation over trivial and
non-trivial (black-hole) space-time topologies in the path over histories
makes an asymptotic observer \textquotedblleft unsure\textquotedblright\ as to
the existence of the microscopic black hole fluctuation thus resulting in no
loss of quantum coherence. However, this sort of argument is plagued not only
by the Euclidean formalism, with its concomitant problems of analytic
continuation, but also by a lack of a concrete proof, at least up to now.
Hence the hypothesis of non-unitary evolution deserves further investigation.

We shall consider \ two approaches to space-time which represent distinct
ontologies. The first, D-particle foam, depends on a picture and partial
description in terms of capture and emission of stringy matter by D-particles
\cite{kmw} based on non-critical strings. At late times after this process the
induced space-time metric has interesting stochastic off-diagonal structure
\cite{sarkar} due to D-particle recoil which has an effect on correlated meson
flavour pairs compatible with the earlier conjectured $\omega$ effect
\cite{bernabeu1},\cite{bernabeu}.The second picture of space-time foam
\cite{garay} that we will consider is based on a heuristic non-local effective
theory approach to gravitational fluctuations. The non-locality arises because
the the fluctuation scale is taken to be intermediate between the Planck scale
and the low energy scale. Furthermore the dominant non-locality was assumed to
be bilocal, and, from similarities to quantum Brownian motion \cite{gardiner}%
,\cite{giulini} plausible arguments can be given for a thermal bath model for
space-time foam \cite{garay}.

It is rare to have a test which can qualitativley distinguish space-time foams
and we will show that the study of correlations in neutral flavoured mesons
can provide one. Earlier work has suggested that this may be the case
\cite{bernabeu} but the analysis for the thermal bath case although suggestive
was quite incomplete. In this paper we will show that the two foams have
qualitatively different behaviour even allowing for non-thermal states of the
bath. Neutral mesons such as the $K$ mesons have in the past been pivotal in
the study of discrete symmetries \cite{wu}.The decay of a (generic) meson
(e.g. the $\phi$ meson) with quantum numbers $J^{PC}=1^{--}$ \cite{lipkin},
leads to a pair state $\left\vert i\right\rangle $ of neutral mesons ($M$)
which has the form of the entangled state%
\begin{equation}
\left\vert i\right\rangle =\frac{1}{\sqrt{2}}\left(  \left\vert \overline
{M_{0}}\left(  \overrightarrow{k}\right)  \right\rangle \left\vert
M_{0}\left(  -\overrightarrow{k}\right)  \right\rangle -\left\vert
M_{0}\left(  \overrightarrow{k}\right)  \right\rangle \left\vert
\overline{M_{0}}\left(  -\overrightarrow{k}\right)  \right\rangle \right)  .
\end{equation}
This state has $CP=+$. If CPT is not defined then $M_{0}$ and $\overline
{M_{0}}$ may not be identified and the requirement of $CP=+$ can be relaxed
\cite{bernabeu1},\cite{bernabeu}. Consequently the state of the meson pair can
be parametrised to have the form
\begin{align}
\left\vert i\right\rangle  &  =\left(  \left\vert \overline{M_{0}}\left(
\overrightarrow{k}\right)  \right\rangle \left\vert M_{0}\left(
-\overrightarrow{k}\right)  \right\rangle -\left\vert M_{0}\left(
\overrightarrow{k}\right)  \right\rangle \left\vert \overline{M_{0}}\left(
-\overrightarrow{k}\right)  \right\rangle \right) \nonumber\\
&  +\omega\left(  \left\vert \overline{M_{0}}\left(  \overrightarrow
{k}\right)  \right\rangle \left\vert M_{0}\left(  -\overrightarrow{k}\right)
\right\rangle +\left\vert M_{0}\left(  \overrightarrow{k}\right)
\right\rangle \left\vert \overline{M_{0}}\left(  -\overrightarrow{k}\right)
\right\rangle \right)  \label{omega}%
\end{align}
where $\omega=\left\vert \omega\right\vert e^{i\Omega}$ is a complex CPT
violating (CPTV) parameter \cite{bernabeu1}. It is useful at this stage to
rewrite the state $\left\vert i\right\rangle $ in terms of the mass
eigenstates. To be specific, from now on we shall restrict ourselves to the
neutral Kaon system $K_{0}-{\overline{K}}_{0}$, which is produced by a $\phi
$-meson at rest, i.e. $K_{0}-{\overline{K}}_{0}$ in their C.M. frame. The CP
eigenstates (on choosing a suitable phase convention for the states
$\left\vert K_{0}\right\rangle $ \ and $\left\vert \overline{K_{0}%
}\right\rangle $ ) are, in standard notation, $\left\vert K_{\pm}\right\rangle
$ with
\begin{equation}
\left\vert K_{\pm}\right\rangle =\frac{1}{\sqrt{2}}\left(  \left\vert
K_{0}\right\rangle \pm\left\vert \overline{K_{0}}\right\rangle \right)  .
\label{cp}%
\end{equation}
The mass eigensates $\left\vert K_{S}\right\rangle $ and $\left\vert
K_{L}\right\rangle $ are written in terms of $\left\vert K_{\pm}\right\rangle
$ as%

\begin{equation}
\left|  K_{L}\right\rangle =\frac{1}{\sqrt{1+\left|  \varepsilon_{2}\right|
^{2}}}\left[  \left|  K_{-}\right\rangle \,+\varepsilon_{2}\left|
K_{+}\right\rangle \right]  \label{Klong}%
\end{equation}

and
\begin{equation}
\left\vert K_{S}\right\rangle =\frac{1}{\sqrt{1+\left\vert \varepsilon
_{1}\right\vert ^{2}}}\left[  \left\vert K_{+}\right\rangle \,+\varepsilon
_{1}\left\vert K_{-}\right\rangle \right]  \label{Kshort}%
\end{equation}
where $\varepsilon_{i},i=1,2$ measure the degree of CP symmetry breaking.

In terms of the mass eigenstates
\begin{equation}
\left\vert i\right\rangle \simeq\mathcal{C}\left\{
\begin{array}
[c]{c}%
\left(  \left\vert K_{L}\left(  \overrightarrow{k}\right)  \right\rangle
\left\vert K_{S}\left(  -\overrightarrow{k}\right)  \right\rangle -\left\vert
K_{S}\left(  \overrightarrow{k}\right)  \right\rangle \left\vert K_{L}\left(
-\overrightarrow{k}\right)  \right\rangle \right)  +\\
\omega\left(  \left\vert K_{S}\left(  \overrightarrow{k}\right)  \right\rangle
\left\vert K_{S}\left(  -\overrightarrow{k}\right)  \right\rangle -\left\vert
K_{L}\left(  \overrightarrow{k}\right)  \right\rangle \left\vert K_{L}\left(
-\overrightarrow{k}\right)  \right\rangle \right)
\end{array}
\right\}  \label{CPTV}%
\end{equation}
where $\mathcal{C=}\frac{\sqrt{\left(  1+\left\vert \varepsilon_{1}\right\vert
^{2}\right)  \left(  1+\left\vert \varepsilon_{2}\right\vert ^{2}\right)  }%
}{\sqrt{2}\left(  1-\varepsilon_{1}\varepsilon_{2}\right)  }$ \cite{bernabeu1}
since the $\varepsilon_{i}$ are small. The states $\left\vert K_{L}\left(
-\overrightarrow{k}\right)  \right\rangle $ and $\left\vert K_{S}\left(
-\overrightarrow{k}\right)  \right\rangle $ can be considered to form a two
level system with
\begin{align}
\left\vert K_{L}\right\rangle  &  =\left\vert \uparrow\right\rangle
\label{Two level}\\
\left\vert K_{S}\right\rangle  &  =\left\vert \downarrow\right\rangle
.\nonumber
\end{align}
Similarly $\left\vert K_{L}\left(  \overrightarrow{k}\right)  \right\rangle $
and $\left\vert K_{S}\left(  \overrightarrow{k}\right)  \right\rangle $ form
another two level system. For convenience we will refer to the $\left\vert
\uparrow\right\rangle $ \ and $\left\vert \downarrow\right\rangle $ as flavour
in the rest of this article. These two two-level systems will be labelled as
$\left\vert \uparrow^{\left(  i\right)  }\right\rangle $ and $\left\vert
\downarrow^{\left(  i\right)  }\right\rangle $ with $i=1,2$. The unnormalised
state \ $\left\vert i\right\rangle $ will then be an example of a state
\begin{equation}
\left\vert \psi\right\rangle =\left\vert \uparrow^{\left(  1\right)
}\right\rangle \left\vert \downarrow^{\left(  2\right)  }\right\rangle
-\left\vert \downarrow^{\left(  1\right)  }\right\rangle \left\vert
\uparrow^{\left(  2\right)  }\right\rangle +\xi\left\vert \uparrow^{\left(
1\right)  }\right\rangle \left\vert \uparrow^{\left(  2\right)  }\right\rangle
+\xi^{\prime}\left\vert \downarrow^{\left(  1\right)  }\right\rangle
\left\vert \downarrow^{\left(  2\right)  }\right\rangle . \label{initstate}%
\end{equation}

In section 2 we will give some details for the D-particle foam and a
perturbative analysis of the gravitationally dressed states. In section 3
similar details will be given for a model of thermal-like foam. We will
demonstrate a qualitative difference between the two foams. For the thermal
case the perturbative analysis is also backed up by a non-perturbative
analysis in order to rule out any possible higher order $\omega$-effect.

\bigskip

\section{Gravitational Decoherence from D-particle Foam}

The most mathematically-consistent model of quantum gravity to date appears to
be \emph{String Theory}, although other approaches, such as Loop quantum
gravity (based on a canonical formalism of quantization) are making progress
~\cite{smolin}.The non-critical (Liouville) string~\cite{ddk} provides a
formalism for dealing with decoherent quantum space-time foam backgrounds,
that include microscopic quantum-fluctuating black holes~\cite{emn}.Given the
limited understanding of gravity at the quantum level, the analysis of
modifications of the quantum Liouville equation implied by non-critical
strings can only be approximate and should be regarded as circumstantial
evidence in favour of a dissipative master equation governing evolution. In
the context of two-dimensional toy black holes \cite{2dbhstring} and in the
presence of singular space-time fluctuations there are believed to be
inherently unobservable delocalised modes which fail to decouple from light
(i.e. the observed) states. The effective theory of the light states which are
measured by local scattering experiments can be described by a non-critical
Liouville string. This results in an irreversible temporal evolution in target
space with decoherence and associated entropy production.

The following master equation for the evolution of stringy low-energy matter
in a non-conformal $\sigma$-model~can be derived \cite{emn}
\begin{equation}
\partial_{t}\rho=i\left[  \rho,H\right]  +:\beta^{i}\mathcal{G}_{ij}\left[
g^{j},\rho\right]  : \label{master}%
\end{equation}
where $t$ denotes time (Liouville zero mode), the $H$ is the effective
low-energy matter Hamiltonian, $g^{i}$ are the quantum background target space
fields, $\beta^{i}$ are the corresponding renormalization group $\beta$
functions for scaling under Liouville dressings and $\mathcal{G}_{ij}$ is the
Zamolodchikov metric \cite{zam,kutasov} in the moduli space of the string. The
double colon symbol in (\ref{master}) represents the operator ordering
$:AB:=\left[  A,B\right]  $ . The index $i$ labels the different background
fields as well as space-time. Hence the summation over $i,j$ in (\ref{master})
corresponds to a discrete summation as well as a covariant integration $\int
d^{D+1}y\,\sqrt{-g}$\bigskip\ where $y$ denotes a set of $\left(  D+1\right)
$-dimensional target space-time co-ordinates and $D$ is the space-time
dimensionality of the original non-critical string.The discovery of new
solitonic structures in superstring theory~\cite{polch} has dramatically
changed the understanding of target space structure. These new
non-perturbative objects are known as D-branes and their inclusion leads to a
scattering picture of space-time fluctuations. The study of D-brane dynamics
has been made possible by Polchinski's realization~\cite{polch} that such
solitonic string backgrounds can be described in a conformally invariant way
in terms of world sheets with boundaries. On these boundaries Dirichlet
boundary conditions for the collective target-space coordinates of the soliton
are imposed.
\begin{figure}[th]
\centering
\includegraphics[
height=2.1551in,
width=2.0989in]{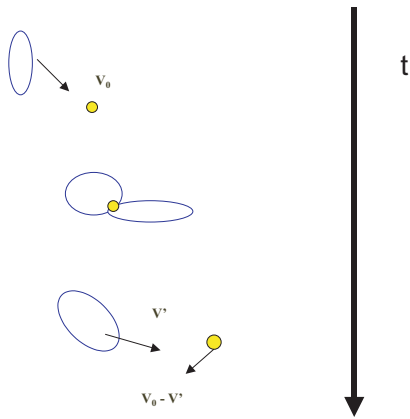}%
\caption{Illustration of approach towards and capture of a closed string
matter state by a D-particle. Subsequently a closed string is emitted by the
D-particle.}%
\label{fig:recoil}%
\end{figure}

Heuristically, when low energy matter given by a closed (or open) string
propagating in a $\left(  D+1\right)  $-dimensional space-time collides with a
very massive D-particle embedded in this space-time, the D-particle recoils as
a result (c.f. Fig. \ref{fig:recoil}). Since there are no rigid bodies in
general relativity the recoil fluctuations of the brane and their effectively
stochastic back-reaction on space-time cannot be neglected.

A concrete model in the form of a supersymmetric space-time foam has been
suggested in \cite{emw} . It is based on parallel brane worlds (with three
large spatial dimensions), moving in a bulk space-time which contains a
\textquotedblleft gas\textquotedblright\ of D-particles. The number of
parallel branes used is dictated by the requirements of target-space
supersymmetry in the limit of zero-velocity branes. One of these branes
represents allegedly our Observable Universe. As the brane moves in the bulk
space, D-particles cross the brane in a random way. From the point of view of
an observer in the brane the crossing D-particles will appear as space-time
defects which flash on and off , i.e. microscopic space-time fluctuations.
This will give the four-dimensional brane world a \textquotedblleft
D-foamy\textquotedblright\ structure.Using this model for space-time
fluctuations we will obtain an expression for the \textit{induced} space-time
distortion as a result of D-particle recoil. In the weakly coupled string
limit, using logarithmic conformal field theory, it can be shown that
\cite{kmw}:
\begin{equation}
g_{mn}=\delta_{mn},\,g_{00}=-1,g_{0n}=\varepsilon\left(  \varepsilon
y_{n}+v_{n}t\right)  \Theta_{\varepsilon}\left(  t\right)  ,\;m,n=1,\ldots,D
\label{recmetr}%
\end{equation}
where the suffix $0$ denotes temporal (Liouville) components and
$\Theta_{\varepsilon}\left(  t\right)  $ is a regularised Heavyside function
with integral representation
\begin{align}
\Theta_{\varepsilon}\left(  t\right)   &  =\frac{1}{2\pi i}\int_{-\infty
}^{\infty}\frac{dq}{q-i\varepsilon}e^{iqt},\label{heaviside}\\
v_{n}  &  =\left(  k_{0}-k^{\prime}\right)  _{n}\;,\nonumber
\end{align}
with $k_{0}\left(  k^{\prime}\right)  $ the momentum of the propagating
closed-string state before (after) the recoil; $y_{n}$ are the spatial
collective coordinates of the D particle and $\varepsilon^{-2}$ is identified
with the target Minkowski time $t$ for $t\gg0$ after the collision~\cite{kmw}
( in units where the string slope $\alpha^{\prime}$ is taken to be $1$). These
relations have been calculated for non-relativistic branes where $v_{n}$ is
small. To leading order for large $t$,%

\begin{equation}
g_{0n}\simeq\overline{v}_{n}\equiv\frac{v_{n}}{\varepsilon}\propto\frac{\Delta
k_{n}}{M_{P}} \label{recoil}%
\end{equation}
where $\Delta k_{n}$ is the momentum transfer during a collision and $M_{P}$
is the Planck mass (actually, to be more precise, $M_{P}=M_{s}/g_{s}$, where
$g_{s}<1$ is the (weak) string coupling, and $M_{s}$ is a string mass scale);
so $g_{0i}$ is constant in space-time but depends on the energy content of the
low energy particle. The operator $\Delta k_{n}$ is complicated to treat but
on considering many collisions it can be regarded, in some sense, as random;
consequently in \cite{sarkar} this transfer process was modelled by a
classical Gaussian random variable $r$ (for an isotropic foam) which
multiplies the momentum operator $\widehat{p}$ for the particle:
\begin{equation}
\overline{u_{i}}\qquad\rightarrow\qquad\frac{r}{M_{P}}\widehat{p}%
\end{equation}
Moreover the mean and variance of $r$ are given by
\begin{equation}
\left\langle r\right\rangle =0~,\qquad\mathrm{and}\qquad\left\langle
r^{2}\right\rangle =\sigma^{2}~.
\end{equation}
The process of capture and emission does not have to conserve flavour.
Consequently we need to generalise the stochastic structure to allow for
this.The fluctuations of each component of the metric tensor $g^{\alpha\beta}$
will then not be just given by the simple recoil distortion (\ref{recoil}),
but instead will be taken to have a $2\times2$ (\textquotedblleft
flavour\textquotedblright) structure \cite{bernabeu}:
\begin{align}
g^{00}  &  =\left(  -1+r_{4}\right)  \mathsf{1}\nonumber\\
g^{01}  &  =g^{10}=r_{0}\mathsf{1}+r_{1}\sigma_{1}+r_{2}\sigma_{2}+r_{3}%
\sigma_{3}\label{stochastic metric}\\
g^{11}  &  =\left(  1+r_{5}\right)  \mathsf{1}\nonumber
\end{align}
where $\mathsf{1}$ , is the identity and $\sigma_{i}$ are\ the Pauli matrices.
The above parametrisation has been taken for simplicity and we will also
consider motion to be in the $x$- direction which is natural since the meson
pairs move collinearly. In any given realisation of the random variables for a
gravitational background,\ the system evolution can be considered to be given
by the Klein-Gordon equation for a two component spinless neutral meson field
$\Phi=\left(
\begin{array}
[c]{c}%
\phi_{1}\\
\phi_{2}%
\end{array}
\right)  $ (corresponding to the two flavours) with mass matrix $m=\frac{1}%
{2}\left(  m_{1}+m_{2}\right)  \mathsf{1}+$ $\frac{1}{2}\left(  m_{1}%
-m_{2}\right)  \sigma_{3}$. We thus have
\begin{equation}
(g^{\alpha\beta}D_{\alpha}D_{\beta}-m^{2})\Phi=0 \label{KleinGordon}%
\end{equation}
where $D_{\alpha}$ is the covariant derivative. Since the Christoffel symbols
vanish (as the $a_{i}$ are independent of space-time) the $D_{\alpha}$
coincide with $\partial_{\alpha}$. Hence within this flavour changing
background (\ref{KleinGordon}) becomes
\begin{equation}
\left(  g^{00}\partial_{0}^{2}+2g^{01}\partial_{0}\partial_{1}+g^{11}%
\partial_{1}^{2}\right)  \Phi-m^{2}\Phi=0. \label{KG2}%
\end{equation}
We shall consider the two particle tensor product states made from the single
particle states $\left\vert \uparrow^{\left(  i\right)  }\right\rangle $ and
$\left\vert \downarrow^{\left(  i\right)  }\right\rangle $ $i=1,2$. This
includes the correlated states which include the state of the $\omega$ effect.
In view of (\ref{KG2}) the evolution of this state is governed by a
hamiltonian $\widehat{H}$
\begin{equation}
\widehat{H}=g^{01}\left(  g^{00}\right)  ^{-1}\widehat{k}-\left(
g^{00}\right)  ^{-1}\sqrt{\left(  g^{01}\right)  ^{2}{k}^{2}-g^{00}\left(
g^{11}k^{2}+m^{2}\right)  } \label{GenKG}%
\end{equation}
which is the natural generalisation of the standard Klein Gordon hamiltonian
in a one particle situation where $\widehat{k}\left\vert \pm k,\alpha
\right\rangle =\pm k\left\vert \pm k,\alpha\right\rangle $ with $\alpha
=\uparrow$ or $\downarrow$. $\widehat{H}$ is a single particle hamiltonian and
in order to study two particle states associated with $i=1,2$ we can define
$\widehat{H}_{i}$ in the natural way (in terms of $\widehat{H}$) and then the
total hamiltonian is $\mathcal{H}=\sum_{i=1}^{2}$ $\widehat{H}_{i}$. The
effect of space-time foam on the initial entangled state of two neutral mesons
is conceptually difficult to isolate, given that the meson state is itself
entangled with the bath. Nevertheless, in the context of our specific model,
which is written as a stochastic hamiltonian, one can estimate the order of
the associated $\omega$-effect of \cite{bernabeu} by applying non-degenerate
perturbation theory to the states $\left\vert \uparrow^{\left(  i\right)
}\right\rangle $, $\left\vert \downarrow^{\left(  i\right)  }\right\rangle $,
$i=1,2$ (where the label $\pm k$ is redundant since $i$ already determines
it). Although it would be more rigorous to consider the corresponding density
matrices, traced over the unobserved gravitational degrees of freedom, in
order to obtain estimates it will suffice formally to work with single-meson
state vectors) .

Owing to the form of the hamiltonian (\ref{GenKG}) the gravitationally
perturbed states will still be momentum eigenstates. The dominant features of
a possible $\omega$-effect can be seen from a term $\widehat{H_{I}}$ in the
interaction hamiltonian
\begin{equation}
\widehat{H_{I}}=-\left(  {r_{1}\sigma_{1}+r_{2}\sigma_{2}}\right)  \widehat{k}
\label{inthamil}%
\end{equation}
which is the leading order contribution in the small parameters $r_{i}$ (c.f.
(\ref{stochastic metric}),(\ref{GenKG})) in $H$ (i.e $\sqrt{\Delta_{i}}$ are
small). In first order in perturbation theory the gravitational dressing of
$\left\vert {\downarrow}^{\left(  i\right)  }\right\rangle $ leads to a state:%

\begin{equation}
\left\vert \downarrow^{\left(  i\right)  }\right\rangle _{QG}=\left\vert
\downarrow^{\left(  i\right)  }\right\rangle +\left\vert \uparrow^{\left(
i\right)  }\right\rangle \alpha^{\left(  i\right)  }%
\end{equation}
where
\begin{equation}
\alpha^{\left(  i\right)  }=\frac{\left\langle \uparrow^{\left(  i\right)
}\right\vert \widehat{H_{I}}\left\vert \downarrow^{\left(  i\right)
}\right\rangle }{E_{2}-E_{1}} \label{qgpert}%
\end{equation}
and correspondingly for $\left\vert {\uparrow}^{\left(  i\right)
}\right\rangle $ the dressed state is obtained from (\ref{qgpert}) by
$\left\vert \downarrow\right\rangle \leftrightarrow\left\vert \uparrow
\right\rangle $ and $\alpha\rightarrow\beta$ where
\begin{equation}
\beta^{\left(  i\right)  }=\frac{\left\langle \downarrow^{\left(  i\right)
}\right\vert \widehat{H_{I}}\left\vert \uparrow^{\left(  i\right)
}\right\rangle }{E_{1}-E_{2}} \label{qgpert2}%
\end{equation}
Here the quantities $E_{i}=(m_{i}^{2}+k^{2})^{1/2}$ denote the energy
eigenvalues, and $i=1$ is associated with the up state and $i=2$ with the down
state. With this in mind the totally antisymmetric \textquotedblleft
gravitationally-dressed\textquotedblright\ state can be expressed in terms of
the unperturbed single-particle states as:%

\[%
\begin{array}
[c]{l}%
\left\vert {\uparrow}^{\left(  1\right)  }\right\rangle _{QG}\left\vert
{\downarrow}^{\left(  2\right)  }\right\rangle _{QG}-\left\vert {\downarrow
}^{\left(  1\right)  }\right\rangle _{QG}\left\vert {\uparrow}^{\left(
2\right)  }\right\rangle _{QG}=\\
\left\vert {\uparrow}^{\left(  1\right)  }\right\rangle \left\vert
{-k,\downarrow}\right\rangle ^{\left(  2\right)  }-\left\vert {k,\downarrow
}\right\rangle ^{\left(  1\right)  }\left\vert {-k,\uparrow}\right\rangle
^{\left(  2\right)  }\\
+\left\vert {\downarrow}^{\left(  1\right)  }\right\rangle \left\vert
{\downarrow}^{\left(  2\right)  }\right\rangle \left(  {\beta^{\left(
1\right)  }-\beta^{\left(  2\right)  }}\right)  +\left\vert {\uparrow
}^{\left(  1\right)  }\right\rangle \left\vert {\uparrow}^{\left(  2\right)
}\right\rangle \left(  {\alpha^{\left(  2\right)  }-\alpha^{\left(  1\right)
}}\right) \\
+\beta^{\left(  1\right)  }\alpha^{\left(  2\right)  }\left\vert {\downarrow
}^{\left(  1\right)  }\right\rangle \left\vert {\uparrow}^{\left(  2\right)
}\right\rangle -\alpha^{\left(  1\right)  }\beta^{\left(  2\right)
}\left\vert {\uparrow}^{\left(  1\right)  }\right\rangle \left\vert
{\downarrow}^{\left(  2\right)  }\right\rangle \\
\label{entangl}%
\end{array}
\]
It should be noted that for $r_{i}\propto\delta_{i1}$ the generated $\omega
$-like effect corresponds to the case $\xi=\xi^{\prime}$ in (\ref{initstate})
since $\alpha^{\left(  i\right)  }=-\beta^{\left(  i\right)  }$, while the
$\omega$-effect of \cite{bernabeu1} (\ref{CPTV}) corresponds to $r_{i}%
\propto\delta_{i2}$ (and the generation of $\xi=-\xi^{\prime}$) since
$\alpha^{\left(  i\right)  }=\beta^{\left(  i\right)  }.$ In the density
matrix these cases can be distinguished by the off-diagonal terms.

On averaging the density matrix over the random variables $r_{i}$, we observe
that only terms of order $|\omega|^{2}$ will survive, with the order of
$|\omega|^{2}$ being
\begin{equation}
|\omega|^{2} = \mathcal{O}\left(  \frac{1}{(E_{1} - E_{2})} (\langle
\downarrow, k |H_{I} |k, \uparrow\rangle)^{2} \right)  = \mathcal{O}\left(
\frac{\Delta_{2} k^{2}}{(E_{1} - E_{2})^{2}} \right)  \sim\frac{\Delta_{2}
k^{2}}{(m_{1} - m_{2})^{2}} \label{omegaorder}%
\end{equation}
for the physically interesting case in which the momenta are of order of the
rest energies.

Recalling (c.f. (\ref{recoil})) that the variance $\Delta_{1}$ is of the order
of the square of the momentum transfer (in units of the Planck mass scale
$M_{P}$) during the scattering of the single particle state off a
space-time-foam defect, i.e. $\Delta_{1}=\zeta^{2}k^{2}/M_{P}^{2}$, where
$\zeta$ is at present a phenomenological parameter. It cannot be further
determined due to the lack of a complete theory of quantum gravity, which
would in principle determine the order of the momentum transfer. We arrive at
the following estimate of the order of $\omega$ in this model of foam:
\begin{equation}
|\omega|^{2}\sim\frac{\zeta^{2}k^{4}}{M_{P}^{2}(m_{1}-m_{2})^{2}}
\label{order}%
\end{equation}
Consequently for neutral kaons, with momenta of the order of the rest energies
$|\omega|\sim10^{-4}|\xi|$, whilst for $B$-mesons we have $|\omega|\sim
10^{-6}|\zeta|$. For $1>\zeta\geq10^{-2}$ these values for $\omega$ are not
far below the sensitivity of current facilities, such DA$\Phi$NE, and $\zeta$
may be constrained by future data owing to possible improvements of the
DA$\Phi$NE detector or a B-meson factory. If the universality of quantum
gravity is assumed then $\zeta$ can also be restricted by data from other
sensitive probes, such as terrestrial and extraterrestrial
neutrinos~\cite{sarkar1}.

\bigskip

\bigskip

\section{Thermal bath model for foam}

\bigskip

Garay \cite{garay} has argued that the effect of non-trivial topologies
related to space-time foam and a non-zero minimum length can be modelled by a
field theory with non-local interactions on a flat background in terms of a
complete set of local functions $\left\{  h_{j}\left(  \phi,x\right)
\right\}  $ of the fields $\phi$ at a space-time point $x$. His argument is
based on general arguments related to problems of measurement \cite{giulini}.
Also, by considering the infinite redhshifting near the horizon for an
observer far away from the horizon of a black hole, Padmanabhan
\cite{padmanabhan} has argued that a foam consisting of virtual black holes
would magnify Planck scale physics for observers asymptotically far from the
horizon and thus an effective non-local field theory description would be
appropriate. The non-local action $I$ can be written in terms of a sum of
non-local terms $I_{n}$ where
\begin{equation}
I_{n}=\frac{1}{n!}\int dx_{1}\ldots\int dx_{n}\,\,f^{i_{1}\ldots i_{n}}\left(
x_{1},\ldots,x_{n}\right)  h_{i_{1}}\left(  x_{1}\right)  h_{i_{2}}\left(
x_{2}\right)  \ldots h_{i_{n}}\left(  x_{n}\right)  \label{nonlocal}%
\end{equation}
(where a summation convention for the indices has been assumed; the
$f^{i_{1}\ldots i_{n}}$ depend on relative co-ordinates such as $x_{1}-x_{2}$
and are expected to fall off for large separations, the scale of fluctuations
being $l_{\ast}>\frac{1}{M_{P}}$ ). Assuming a form of weak coupling
approximation it was further argued \cite{garay} that the retention of only
the $n=2$ term would be a reasonable approximation. Formally (i.e. ignoring
the validity of Euclidean to Minkowski Wick rotations) the resulting
non-locality can be written in terms of an auxiliary field $\varphi$ through
the functional identity
\begin{align}
&  \exp\left(  i\int dx_{1}\int dx_{2}\,\,f^{i_{1}i_{2}}\left(  x_{1}%
-x_{2}\right)  h_{i_{1}}\left(  x_{1}\right)  h_{i_{2}}\left(  x_{2}\right)
\right) \nonumber\\
&  =\int d\varphi\exp\left(  -\int\int dx_{1}dx_{2}\,k_{i_{1}i_{2}}\left(
x_{1}-x_{2}\right)  \varphi^{i_{1}}\left(  x_{1}\right)  \varphi^{i_{1}%
}\left(  x_{2}\right)  \right) \\
&  \times\exp\left(  i\int dx\,\varphi^{j}\left(  x\right)  h_{j}\left(
x\right)  \right) \nonumber
\end{align}
where $k_{i_{1}i_{2}}$ is the inverse of $f^{i_{1}i_{2}}$. The bilocality is
now represented as a local field theory for $\phi$ subjected to a stochastic
field $\varphi$. As shown in \cite{barenboim2} and \cite{loreti} this
stochastic behaviour results \cite{garay} in a master equation \ of the type
found in quantum Brownian motion, i.e. unitary evolution supplemented by
diffusion. There is no dissipation which might be expected from the
fluctuation diffusion theorem because the noise is classical. \ Since the
energy scales for typical experiments are much smaller than those associated
with gravitational quantum fluctuations Garay considered $f^{i_{1}i_{2}%
}\left(  x_{1}-x_{2}\right)  $ to be proportional to a Dirac delta function
and argued that the master equation was that for a thermal bath.

A thermal field represents a bath about which there is minimal information
since only the mean energy of the bath is known, a situation which maybe valid
also for space-time foam. In applications of quantum information it has been
shown that a system of two qubits (or two-level systems) initially in a
separable state (and interacting with a thermal bath) can actually be
entangled by such a single mode bath \cite{entanglement}. As the system
evolves the degree of entanglement is sensitive to the initial state. The
close analogy between two-level systems and neutral meson systems, together
with the modelling by a phenomenological thermal bath of space-time foam,
makes the study of thermal master equations an intriguing one for the
generation of $\omega$-terms.The hamiltonian $\mathcal{H}$ representing the
interaction of \ two such two-level `atoms' with a single mode thermal field
\cite{jaynes} is
\begin{equation}
\mathcal{H}=\nu a^{\dag}a+\frac{1}{2}\Omega\sigma_{3}^{\left(  1\right)
}+\frac{1}{2}\Omega\sigma_{3}^{\left(  2\right)  }+\gamma\sum_{i=1}^{2}\left(
a\sigma_{+}^{\left(  i\right)  }+a^{\dag}\sigma_{-}^{\left(  i\right)
}\right)  \label{thermal}%
\end{equation}
where $a$ is the annihilation operator for the mode of the thermal field and
the $\sigma$'s are again the Pauli matrices for the 2-level systems (using the
standard conventions).The harmonic oscillator operators $a$ and $a^{\dag}$
satisfy
\begin{equation}
\left[  a,a^{\dag}\right]  =1,\left[  a^{\dag},a^{\dag}\right]  =\left[
a,a\right]  =0. \label{commutation}%
\end{equation}
The $\mathcal{H}$ here is quite different from the $\mathcal{H}$ of the
D-particle foam of the last section. There are no classical stochastic terms
at this level of description and also $\mathcal{H}$ is not separable. In the
D-particle foam model the lack of separabillity came solely from the entangled
nature of the initial unperturbed state. The thermal master equation comes
from tracing over the oscillator degrees of freedom. We will however consider
the dynamics before tracing over the bath because, although the thermal bath
idea has a certain intuitive appeal, it cannot claim to be rigorous and so for
attempts to find models for the $\omega$-effect it behoves us to entertain
also deviations from the thermal bath state of the reservoir.

An important feature of \ $\mathcal{H}$ in \ref{thermal} is the block
structure of subspaces that are left invariant by $\mathcal{H}$. It is
straightforward to show that the family of invariant irreducible spaces
$\mathcal{E}_{n}$ may be defined by $\left\{  \left\vert e_{i}^{\left(
n\right)  }\right\rangle ,i=1,\ldots,4\right\}  $ where ( in obvious notation,
with $n$ denoting the number of oscillator quanta)
\begin{align}
\left\vert e_{1}^{\left(  n\right)  }\right\rangle  &  \equiv\left\vert
\uparrow^{\left(  1\right)  },\,\uparrow^{\left(  2\right)  },n\right\rangle
,\label{equation1}\\
\left\vert e_{2}^{\left(  n\right)  }\right\rangle  &  \equiv\left\vert
\uparrow^{\left(  1\right)  },\,\downarrow^{\left(  2\right)  },n\right\rangle
,\nonumber\\
\left\vert e_{3}^{\left(  n\right)  }\right\rangle  &  \equiv\left\vert
\downarrow^{\left(  1\right)  },\,\uparrow^{\left(  2\right)  },n\right\rangle
,\nonumber
\end{align}
and
\[
\left\vert e_{4}^{\left(  n\right)  }\right\rangle \equiv\left\vert
\downarrow^{\left(  1\right)  },\,\downarrow^{\left(  2\right)  }%
,n\right\rangle .
\]
The total space of states is a direct sum of the $\mathcal{E}_{n}$. We will
write $\mathcal{H}$ as $\mathcal{H}_{0}+\mathcal{H}_{1}$ where
\[
\mathcal{H}_{0}=\nu a^{\dag}a+\frac{\Omega}{2}\left(  \sigma_{3}^{\left(
1\right)  }+\sigma_{3}^{\left(  2\right)  }\right)
\]
and
\[
\mathcal{H}_{1}=\gamma\sum_{i=1}^{2}\left(  a\sigma_{+}^{\left(  i\right)
}+a^{\dag}\sigma_{-}^{\left(  i\right)  }\right)  .
\]
$n$ is a quantum number and gives the effect of the random environment. In our
era the strength $\gamma$ of the coupling with the bath is weak. We expect
heavy gravitational degrees of freedom and so $\Omega\gg\nu$. It is possible
to associate both thermal and highly non-classical density matrices with the
bath state. We will investigate whether this freedom is enough to generate
$\omega$-type terms.

\bigskip\ We will calculate the stationary states in $\mathcal{E}_{n}$, using
degenerate perturbation theory, where appropriate. We will be primarily
interested in the dressing of the degenerate states $\left\vert e_{2}^{\left(
n\right)  }\right\rangle $ and $\left\vert e_{3}^{\left(  n\right)
}\right\rangle $ because it is these which contain the neutral meson entangled
state . In 2nd order perturbation theory the dressed states are
\begin{equation}
\left\vert \psi_{2}^{\left(  n\right)  }\right\rangle =\left\vert
e_{2}^{\left(  n\right)  }\right\rangle -\left\vert e_{3}^{\left(  n\right)
}\right\rangle +O\left(  \gamma^{3}\right)  \label{dressed1}%
\end{equation}
with energy $E_{2}^{\left(  n\right)  }=\left(  n+1\right)  \nu+O\left(
\gamma^{3}\right)  $ and
\begin{equation}
\left\vert \psi_{3}^{\left(  n\right)  }\right\rangle =\left\vert
e_{2}^{\left(  n\right)  }\right\rangle +\left\vert e_{3}^{\left(  n\right)
}\right\rangle +O\left(  \gamma^{3}\right)  \label{dressed2}%
\end{equation}
with energy $E_{3}^{\left(  n\right)  }=\left(  n+1\right)  \nu+\frac
{2\gamma^{2}}{\Omega-\nu}+O\left(  \gamma^{3}\right)  $. It is $\left\vert
\psi_{2}^{\left(  n\right)  }\right\rangle $ which can in principle give the
$\omega$-effect. More precisely we would construct the state $Tr\left(
\left\vert \psi_{2}^{\left(  n\right)  }\right\rangle \left\langle \psi
_{2}^{\left(  n\right)  }\right\vert \rho_{B}\right)  $ where $\rho_{B}$ is
the bath density matrix (and has the form $\rho_{B}=\sum_{n,m}p_{nn^{\prime}%
}\left\vert n\right\rangle \left\langle n^{\prime}\right\vert $ for suitable
choices of $p_{nn^{\prime}}$). To this order of approximation $\left\vert
\psi_{2}^{\left(  n\right)  }\right\rangle $ cannot generate the $\omega
$-effect since there is no admixture of $\left\vert e_{1}^{\left(  n\right)
}\right\rangle $ and $\left\vert e_{4}^{\left(  n\right)  }\right\rangle $.
However it is a priori possible that this may change when higher orders in
$\gamma$. We can show that $\left\vert \psi_{2}^{\left(  n\right)
}\right\rangle =\left\vert e_{2}^{\left(  n\right)  }\right\rangle -\left\vert
e_{3}^{\left(  n\right)  }\right\rangle $ to all orders in $\gamma$ by
directly considering the hamiltonian matrix $\mathcal{H}^{\left(  n\right)  }$
for $\mathcal{H}$ within $\mathcal{E}_{n}$; it is given by
\begin{equation}
\mathcal{H}^{\left(  n\right)  }=\left(
\begin{array}
[c]{cccc}%
\Omega+n\nu & \gamma\sqrt{n+1} & \gamma\sqrt{n+1} & 0\\
\gamma\sqrt{n+1} & \left(  n+1\right)  \nu & 0 & \gamma\sqrt{n+2}\\
\gamma\sqrt{n+1} & 0 & \left(  n+1\right)  \nu & \gamma\sqrt{n+2}\\
0 & \gamma\sqrt{n+2} & \gamma\sqrt{n+2} & \left(  n+2\right)  \nu-\Omega
\end{array}
\right)  . \label{matrix}%
\end{equation}
We immediately notice that
\begin{equation}
\mathcal{H}^{\left(  n\right)  }\left(
\begin{array}
[c]{c}%
0\\
1\\
-1\\
0
\end{array}
\right)  =\left(  n+1\right)  \nu\left(
\begin{array}
[c]{c}%
0\\
1\\
-1\\
0
\end{array}
\right)  \label{eigenvectoreqn}%
\end{equation}
and so, to all orders in $\gamma$, the environment does \textit{not} dress the
state of interest to give the $\omega$-effect; clearly this is independent of
any choice of $\rho_{B}$. One cannot say of course that other more complcated
models of `thermal' baths may not display the $\omega$-effect but clearly a
rather standard model rejects quite emphatically the possibility of such an
effect. This by itself is very interesting. \ It shows that the $\omega
$-effect is far from a generic possibility for space-time foams. Just as it is
remarkable that the commonplace `thermal' bath cannot accommodate the $\omega
$-effect, it also remarkable that the D-particle foam model manages to do so
very simply.

\bigskip

\section{Conclusions}

\bigskip The magnitude of the $\omega$-effect which may not be far from the
sensitivity of immediate future experimental facilities, such as a possible
upgrade of the DA$\Phi$NE detector or a B-meson factory. In this work we have
discussed two classes of space-time foam models, which are not inconceivable
that may characterise realistic situations of the (still elusive) theory of
quantum gravity. It is interesting to continue the search for more realistic
models of quantum gravity, either in the context of string theory or in other
approaches, such as the canonical approach or the loop quantum gravity, in
order to search for intrinsic CPT Violating effects in sensitive matter
probes. Moreover the phenomenological approach of foam baths may be guided by
consideration of hydrodynamical theories \cite{anastopoulos}. Detailed
analyses of global data, including very sensitive probes such as high energy
neutrinos, is the only way forward in order to obtain some clues on the
elusive theory of quantum gravity. Decoherence, induced by quantum gravity, is
not only not ruled out at present, but indeed it may provide the link between
theory and experiment in this elusive area of physics.

\bigskip

\begin{acknowledgement}
This work is partially supported by the European Union through the FP6 Marie
Curie Research and Training Network \emph{UniverseNet} (MRTN-CT-2006-035863).
I would like to thank N.E. Mavromatos for stimulating discussions.
\end{acknowledgement}

\end{document}